\ifx\pdfsuppresswarningpagegroup\undefined\else\pdfsuppresswarningpagegroup=1\fi
\PassOptionsToPackage{hyperfootnotes=false}{hyperref}

\documentclass[a4paper,fleqn]{cas-dc}
\usepackage[numbers]{natbib}
\usepackage{graphicx}
\usepackage{xcolor}
\usepackage{listings}
\usepackage{epstopdf}
\usepackage{subfigure}
\usepackage{hyperref}
\hypersetup{colorlinks, citecolor=green, filecolor=black, linkcolor=blue, urlcolor=blue }
\usepackage{graphicx}
\usepackage{amsmath}
\usepackage[version=4]{mhchem}
\usepackage{siunitx}
\usepackage{longtable,tabularx}
\usepackage{array,tabularx}
\usepackage{multirow,booktabs} 
\usepackage{pifont}

\def\tsc#1{\csdef{#1}{\textsc{\lowercase{#1}}\xspace}}
\tsc{WGM}
\tsc{QE}
\tsc{EP}
\tsc{PMS}
\tsc{BEC}
\tsc{DE}

\begin{document}
	\let\WriteBookmarks\relax
	\def\floatpagepagefraction{1}
	\def\textpagefraction{.001}
	
	\shorttitle{Uncertainty Bounds for Multivariate Machine Learning Predictions on High-Strain Brittle Fracture}
	\shortauthors{Cristina Garcia-Cardona et~al.}
	
	\title [mode = title]{Uncertainty Bounds for Multivariate Machine Learning Predictions on High-Strain Brittle Fracture}

	\author[1]{Cristina Garcia-Cardona}[orcid=0000-0002-5641-3491]
	\cormark[1]
	\ead{cgarciac@lanl.gov}
	\credit{Conceptualization, Methodology, Software, Formal analysis, Visualization, Writing - Original Draft, Writing - Review \& Editing}
	
	\author[2]{M. Giselle Fern\'andez-Godino}[orcid=0000-0002-3837-8661]
	\ead{fernandez48@llnl.gov}
	\credit{Conceptualization, Investigation, Software, Formal analysis, Visualization, Writing - Original Draft, Writing - Review \& Editing}
	
	\author[3]{Daniel O'Malley}[orcid=0000-0003-0432-3088]
	\ead{omalled@lanl.gov}
	\credit{Conceptualization, Writing - Review \& Editing}
	
	\author[4]{Tanmoy Bhattacharya}[orcid=0000-0002-1060-652X]
	\ead{tanmoy@lanl.gov}
	\credit{Conceptualization, Methodology, Supervision, \\ Project administration, Writing - Review \& Editing}
	
	\address[1]{MS B256, Information Sciences Group (CCS-3), Los Alamos National Laboratory, Los Alamos, NM 87545}
	\address[2]{L-103, Atmospheric Science Research \& Applications Group, AEED, Lawrence Livermore National Laboratory, 7000 East Ave, Livermore, CA 94550}
	\address[3]{MS D446, Computational Earth Sciences Group (EES-16), Los Alamos National Laboratory, Los Alamos, NM 87545}
	\address[4]{MS B285, Nuclear and Particle Physics, Astrophysics \& Cosmology Group (T-2), Los Alamos National Laboratory, Los Alamos, NM 87545}
	
	\cortext[cor1]{Corresponding author}
	
	\begin{abstract}
		Simulation of the crack network evolution on high strain rate impact experiments performed in brittle materials is very compute-intensive. The cost increases even more if multiple simulations are needed to account for the randomness in crack length, location, and orientation, which is inherently found in real-world materials. Constructing a machine learning emulator can make the process faster by orders of magnitude. There has been little work, however, on assessing the error associated with their predictions. Estimating these errors is imperative for meaningful overall uncertainty quantification. In this work, we extend the heteroscedastic uncertainty estimates to bound a multiple output machine learning emulator. We find that the response prediction is accurate within its predicted errors, but with a somewhat conservative estimate of uncertainty.
	\end{abstract}
		

	\maketitle
	
	\section{Introduction}
	
	Emulators based on machine learning (ML) have proven to be powerful tools for prediction. These techniques can decode complex coupling between variables, producing surrogates that require substantially reduced computational resources. Nevertheless, ML emulators, particularly those built using deep learning, have been strongly criticized because of their ``black box'' character, resulting in their cautious adoption in computational science. This criticism is ameliorated by uncertainty quantification (UQ) methods that can bound the calculation errors. Moreover, in numerous fields, such as medicine or engineering, it is essential to be able to estimate the level of confidence in the predictions of the model~\cite{Begoli2019}. Uncertainty quantification encompasses a series of mathematical techniques that characterize the space of outcomes of a model while considering that not all the parameters are precisely specified. As such, sensitivity analysis, data assimilation, design of experiments, and risk assessment are important applications of UQ~\cite{smith-2013-uq}. Different techniques have been developed to try to overcome the difficulty of assessing uncertainties for complex ML emulators, including dropout~\cite{gal-2016-dropout} and deep ensembles~\cite{lakshmin-2017-deep}, and for a diverse range of applications such as PDEs~\cite{zhang-2019-pinn}, stochastic PDEs~\cite{zhu-2018-spde}, multiscale methods~\cite{chan-2018-multiscale} and computer vision~\cite{kendall-2017-het}. In this work, we extend tools to characterize heteroscedastic uncertainty~\cite{kendall-2017-het} to the analysis of ML emulators with multiple outputs.
	
	ML emulators have been successfully used for applications such as classification~\cite{suthaharan2014big}, regression~\cite{huang2019application}, bridging scales~\cite{raissi2017physics,cheng2019bridging,wang2020deep}, and dimensionality reduction~\cite{cichocki2016tensor} problems. In recent years, there has been substantial growth in ML application for material science~\cite{schmidt2019recent}, in particular for bridging scales in fracture mechanics. One of the major challenges in this context is the discrepancy in scales between microscale cracks and the macroscale associated with bulk materials~\cite{srinivasan2018quantifying}. Recent work has utilized ML algorithms to study brittle material under low--strain-rate tensile dynamic loading~\cite{moore2018predictive,hunter2018reduced,panda2020mesoscale}, including recursive ML prediction for fracture behavior~\cite{schwarzer2019learning}. The evolution, growth, and interaction of cracks is key to modeling damage behavior in several brittle materials such as granite, concrete, metals, and ceramics~\cite{meyer2000crack, paliwal2008interacting, escobedo2014effect, huq2019micromechanics}. A recent novel alternative is to bridge continuum and mesoscales by developing and implementing a continuum-scale effective-moduli constitutive model that is informed by crack statistics generated from the mesoscale simulations in low--strain-rate~\cite{vaughn2019statistically} or high--strain-rate~\cite{larkin2020scale} conditions. However, the cost of generating large datasets can also be prohibitive if this comes from computationally intensive high-fidelity simulations. Previous work~\cite{fernandez2021accelerating} showed that an inexpensive ML emulator can inform effective moduli when trained using damage and stress information from a mesoscale model. The ML emulator can then be combined with a continuum-scale hydrodynamic simulator to make accurate predictions inexpensively. In this work, we propose a technique to assign the ML emulator uncertainty bounds that can quantify its confidence taking into account the training data variability. Since the uncertainty is rather small during most of the evolution, but is large at some specific regions, an overall measure of the expected prediction error is not the goal; rather, we model the emulator as a heteroscedastic process where the error varies depending on the input state.
	
	\section{Problem of Interest: Flyer Plate Problem} \label{sec:problem}
	
	We study the damage behavior of beryllium under dynamic loading conditions. The experiment is based on a flyer disc impact against a target disc specimen (Cady et al.~\cite{cady2011alamos}). The beryllium samples were machined from a vacuum hot-pressed billet of beryllium S200F grade. Table~\ref{table:Beryllium} shows the main details of the beryllium disc composition, machining, and dimensions.
	
	\begin{table}[ht!]
		\centering
		\begin{tabular}{ccc}
			\hline
			Parameter&Value\\
			\hline
			Beryllium grade & S200F\\
			Flyer disc height	 & $2 mm$\\
			Target disc height & $4 mm$\\ 
			Flyer/target disc diameter  & $28.8 mm$\\
			Flyer/target flat within & $2\mu m$\\
			Flyer/target parallel within & $3\mu m$\\
			Density& $1.85 g/cm^3$\\
			Beryllium content&  $0.72 wt.\%$\\
			Average grain size&  $11.4 \mu m$\\
			Polish & $1 \mu m$ diamond paste\\
			Longitudinal wave sound speed  & $[13.19-13.20] mm/ \mu s$\\
			Shear wave sound speed & $[9.04-9.07] mm/ \mu s$\\
			\hline
		\end{tabular}
		\caption{\label{table:Beryllium}Details for the Beryllium experiential samples.}
	\end{table} 
	
\begin{figure}[ht!] 
		\centering
		\includegraphics[width=0.9\columnwidth]{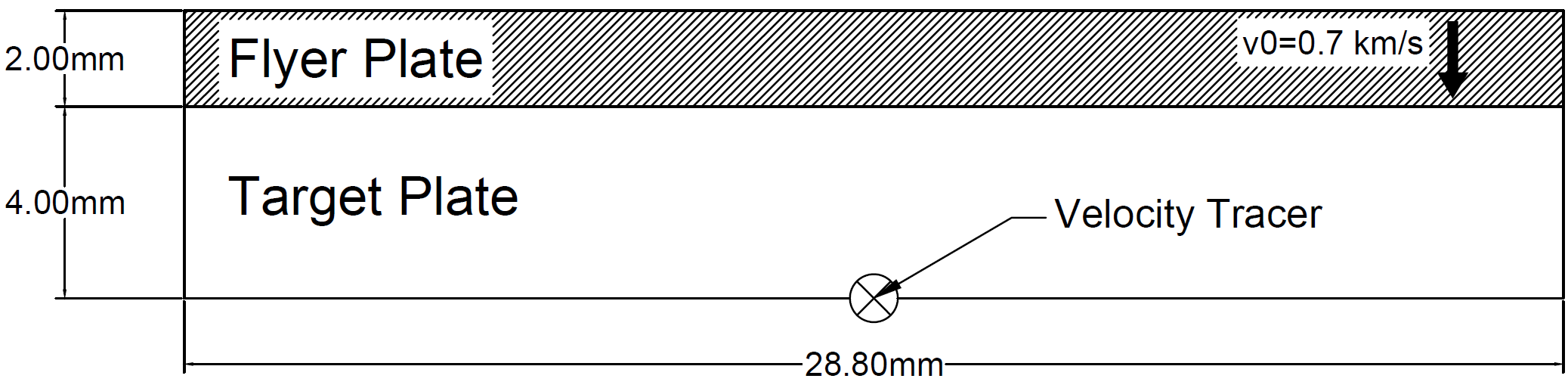} 
		\caption{Initial setup for the flyer plate test simulations. The flyer plate has an initial velocity of $0.721km/s$ and it is initially in contact with the target plate.}
		\label{FigSetup}
	\end{figure}

	For validation purposes, the geometry, the size, the material properties, and the initial loading of the experiments were reproduced in simulations as closely as possible. The simulations were two-dimensional and modeled a beryllium flyer plate impact against a beryllium target specimen. Figure~\ref{FigSetup} is a schematic of the simulation setup. The flyer and target have a width of $28.8mm$, the height of the target is $4 mm$, and the height of the flyer plate is $2mm$. The flyer plate has an initial vertical velocity of 0.721 km/s towards the target plate, and the total simulation time is 1.2$\mu s$. A velocity tracer was placed at the middle rear of the target plate to measure the shock wave profiles, enabling comparison with the experiments and validation. 
	
	After the impact, the target plate is subject to a strong compression that later becomes tension as the shock wave travels within the material and bounces against the borders of the plate. This indirect uniaxial tensile load leads to a Mode I crack growth dominated problem (the loading is applied perpendicular to the crack). The nature of fracture in flyer plate experiments leads to a non-homogeneous damage distribution where a concentrated region of damage forms across the target plate's midspan. In contrast, the majority of the plate remains relatively undamaged. For more information on flyer plate experiments, the reader can refer to the references~\cite{cady2011alamos,cady2012characterization}.

	\subsection*{HOSS Model}\label{sec:HOSS}
	
	In this work, data generated with the Hybrid Optimization Software Suite (HOSS)~\cite{rougier2013lanl,knight2013lanl,knight2015hybrid} was used to build the ML emulator described in Section~\ref{sec:ML}. Modeling samples in HOSS is done by using discrete elements that are further divided into finite elements. The finite-discrete-element method (FDEM) included in HOSS can model the evolution of the microcrack network in high-strain rate problems. The governing equations are conservation of mass, momentum, and energy along with Newton's laws~\cite{munjiza1992discrete, munjiza1995combined, munjiza2004combined, munjiza2011computational, rougier2014validation, munjiza2015large} and time-integration is done using a central difference scheme~\cite{rougier2004numerical}. The cracks are located in the boundary of the finite elements, and often hundreds of elements are needed to model a single fracture~\cite{munjiza2004combined}. The fine grids required, along with the explicit integration scheme, result in very expensive simulations.
	
	HOSS high-fidelity model has been validated against experiments in a number of settings including Split Hopkinson Bar tests on granite~\cite{rougier2014validation}, failure processes in shale~\cite{carey2015fracture}, fracture coalescence processes in granite~\cite{euser2019simulation} and earthquake damage~\cite{klinger2018earthquake}. HOSS can also account for deformation in metals through a recent plasticity model~\cite{rougier2020combined}. HOSS explicitly accounts for crack nucleation, evolution, and coalescence. However, it does not account for microstructure, deformation twinning, dislocations, or atomic breaking at crack tips. The problem of interest in this work can be considered a pure tension problem dominated by opening failure. Still, since not every discrete element edge is oriented orthogonally to the applied load in HOSS, both shear and tearing modes occur at a local mesh element scale. The connections between finite elements are made using springs, so if two elements are pulled apart, a small space appears between them. This also allows one element to slide relative to another.
		
	To recreate the flyer plate simulations in HOSS the inputs needed are the flyer plate mesh, the target plate mesh, the material properties (Beryllium in this case), the distribution and length of the initial cracks in the target plate and the initial velocity of the flyer plate (the target plate is stationary). HOSS is a deterministic model; hence, to obtain the statistical variability naturally existent in materials, we randomly generate the initial crack location, orientation, and length. There are 200 initial cracks, and they are only imposed in the target specimen. A uniform distribution is used to determine the initial crack location $(x,y)$ within the target plate. The horizontal coordinate distribution corresponds to $x \sim U[0,28.8 mm]$ while the vertical coordinate distribution is $y \sim U[0,4 mm]$ (see Figure~\ref{FigSetup}). The initial orientation of the cracks ($\theta$) follows the uniform distribution $\theta \sim U[0^\circ,180^\circ)$. The initial crack lengths are determined based on a power-law probability density function~\cite{bonnet2001scaling, ignatovich2019power}, and the lengths vary between $0.1mm$ and $0.3mm$. The location, length, and orientation distributions generate only the initial conditions for the crack network within the target plate and are changed randomly in every simulation. The probability density function (power-law function) used to generate the initial crack length distribution on the target plate, $f_1(a,t=0)$, is
	\begin{equation}\label{eq1}
	f_1(a,t=0)=\frac{qa^{(q-1)}}{a_2^q-a_1^q},
	\end{equation}
	where $q=-3$, $a_1=0.1$, $a_2=0.3$, and $a$ is a real number in the range $[a_1,a_2]$. The end values $a_1$ and $a_2$ are the initial minimum and maximum crack length. Therefore,
	\begin{equation}\label{eq2}
	f_1(a,t=0)\approx \frac{a^{-4}}{321}.
	\end{equation}
	
\begin{figure}[ht!] 
		\centering
		\includegraphics[width=0.85\columnwidth]{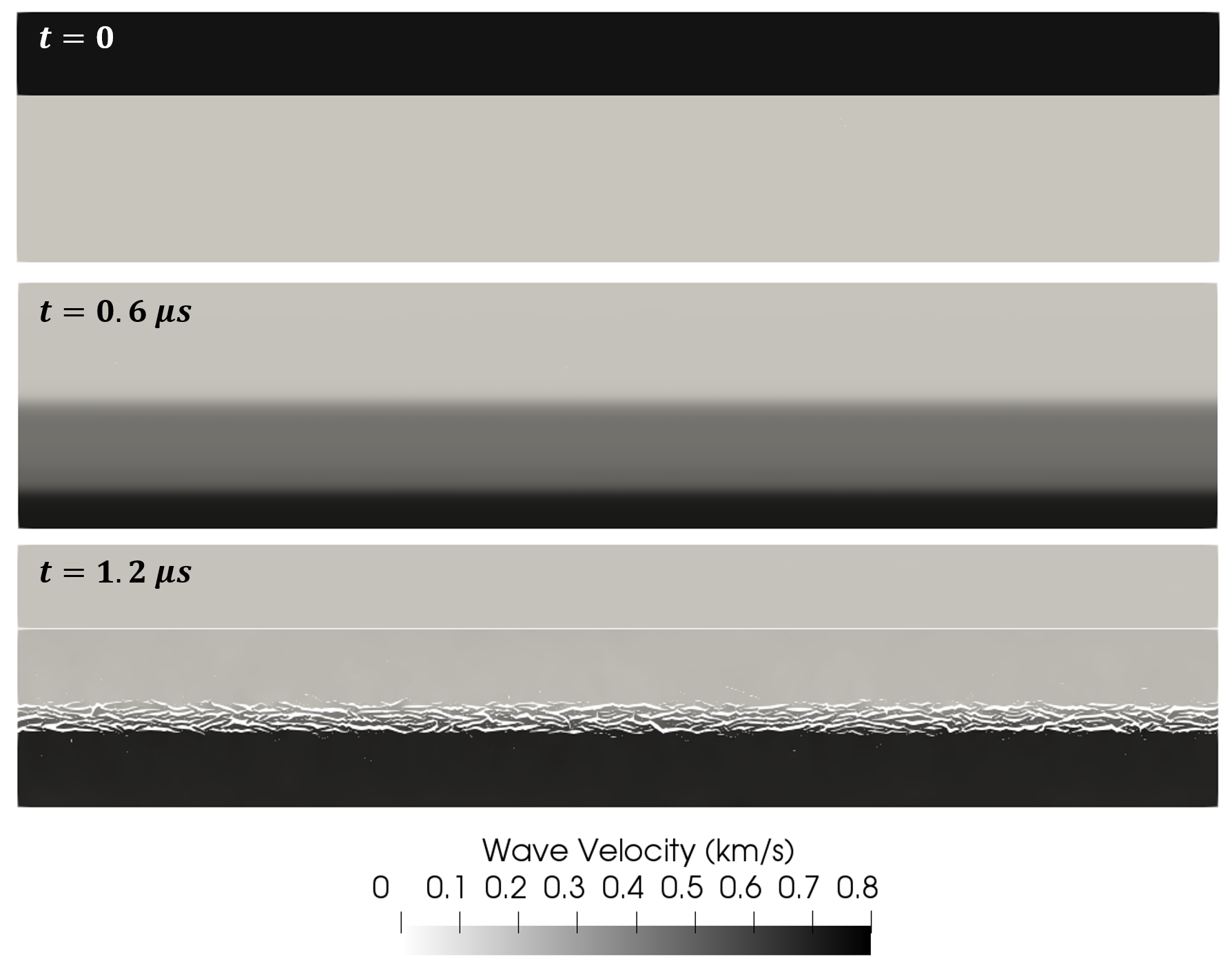} 
		\caption{HOSS simulation. Shock wave velocity at initial, intermediate and final times.}
		\label{fig:velocity}
	\end{figure}

	The evolution of the crack network takes place within the HOSS simulations. The time-dependent crack probability density function is obtained from each HOSS simulation after its completion. Each simulation spans $1.2 \mu s$ using HOSS time steps of $10^{-5} \mu s$, with outputs every $0.0025 \mu s$, which we refer to as a time step---each simulation is thus 480 time-steps long. Figure~\ref{fig:velocity} shows the shock wave velocity at the initial, intermediate and final time ($t=0$, $t=0.6 \mu s$, $t=1.2 \mu s$, respectively) for a HOSS simulation of the flyer plate problem. At $t=1.2 \mu s$, it is observed that the failure is produced in the middle section of the target plate because the target plate is twice the height of the flyer plate (see Figure~\ref{FigSetup}). Note that nucleation and coalescence of cracks occur as the simulation proceeds.
	
	A single simulation of the flyer plate problem, requires 87,066 elements in HOSS to describe the crack dynamics accurately. Each simulation takes $160$-CPU hours and produces 23 GBs of simulation data. Available data for this flyer plate problem includes a total of 100 HOSS simulations. The HOSS computational cost associated is prohibitive for many applications where multiple simulations are needed, such as optimization and UQ, leading to the need for cheaper ML emulators.

	\section{Uncertainty Quantification of Machine Learning Emulator}\label{sec:ML}
	
	The goal of UQ is to assign a level of confidence to ML emulator predictions. In turn, this assignment can be used to estimate decision risks or provide a principled selection mechanism of the most impactful experiments: the ones that minimize the uncertainty. In the case of ML emulators, the UQ approaches are designed to predict both the regular input-output mapping between features and quantities of interest, together with an additional set of outputs intended to capture the level of confidence in the ML emulator predictions. Here we use a heteroscedastic~\cite{kendall-2017-het} set up to represent the uncertainty in the regressor emulator. The heteroscedastic formulation characterizes the output of the regression emulator as a normally-distributed random variable with parameters that are feature-dependent, i.e., the variance of the predictions is heterogeneous and domain-dependent. In contrast with many applications that consist of only one output, we explore the case of UQ for emulators with multiple outputs. Further, we develop a framework for constructing multivariate heteroscedastic UQ bounds. 
	
	\subsection{Machine Learning Emulator}
	
	Since simulating the evolution of the microcrack network in high-strain rate problems is computationally very expensive, ML emulators are gaining traction as alternative data-driven surrogate emulators for this application. With enough data, these emulators can synthesize the system's dynamics, achieving accuracy comparable to the high-fidelity models. At the same time, they are more computationally efficient and require much less time for evaluation. 
	
	In previous work~\cite{fernandez2021accelerating}, we demonstrated that a recurrent neural network (RNN) was able to predict two quantities of interest, namely the length of the longest crack, $L_{\mathrm{long}}$, and the maximum tensile stress, $S_{yy}$, as a function of time for the flyer plate problem described in Section~\ref{sec:problem}. In this work we estimate the uncertainty in the one-time-step response prediction of the coupled system formed by the same quantities of interest: $L_{\mathrm{long}}$ and $S_{yy}$. The more complex task of estimating uncertainties in a multivariate RNN emulator will be addressed in future work.
	
	To build the emulator, the quantities of interest are paired at each time step $t$ as 
	\begin{equation}
	\left ( L_{\mathrm{long}}^{(t)}, S_{yy}^{(t)} \right )\, . \label{eq:y}
	\end{equation}	
	Given a sequence of $d$ consecutive time steps:   
	\begin{equation}
	\left [ \left ( L_{\mathrm{long}}^{(t-d)}, S_{yy}^{(t-d)} \right ), \dots, \left ( L_{\mathrm{long}}^{(t-2)}, S_{yy}^{(t-2)} \right ), \left ( L_{\mathrm{long}}^{(t-1)}, S_{yy}^{(t-1)} \right ) \right ]\, , \label{eq:x}
	\end{equation}
	the emulator is trained to predict the next pair in the time sequence $\left ( L_{\mathrm{long}}^{(t)}, S_{yy}^{(t)} \right )$. Accordingly, the input patterns for the emulator conform to sequences as the one in (\ref{eq:x}), while the output consist of pairs as in (\ref{eq:y}). To simplify the notation the $i$-th input pattern is denoted by $\mathbf{x_i}$ and the corresponding output pair is denoted by $\mathbf{y_i}$. Hence, the training data set is denoted as the collection $\{\mathbf{x_i}, \mathbf{y_i}\}_{i=1}^N$, where $\mathbf{x_i} \in \mathbb{R}^{2d}$, $\mathbf{y_i} \in \mathbb{R}^2$ and $N$ represents the number of samples in the training set.
	
	We design the emulator as a multilayer feed-forward neural network composed of neurons with dense connections. The output $o_j^i$ of each artificial neuron $j$ in layer $i$ is computed as
	\begin{equation} 
	o_j^i = h \left ( \mathbf{w}_{j}^i \cdot \boldsymbol{\xi}^i + b_j^i\right )\, , \label{eq:an}
	\end{equation} 
	where $\boldsymbol{\xi}^i$ represents the input vector at layer $i$; $\mathbf{w}_{j}^i$ and $b_j^i$  represent neuron parameters: weight vector and bias, respectively; the operator $\cdot$ denotes a dot product; and $h$, the activation function. The activation function used is a rectified linear unit (ReLU) and corresponds to the following operation: $\mathrm{ReLU}(\nu) = \max(0, \nu)$. 
	
	In a feed-forward network, the information propagates layer-wise: the input vector $\boldsymbol{\xi}^i$ at layer $i$ is constructed by concatenation of the outputs of the neurons at layer $i-1$, with layer $0$ being the input layer and the last layer being the output layer, whose output constitutes the output of the network. Intermediate layers (i.e., different from the input and output layers) are called hidden layers. The overall mapping computed by the neural network can be denoted as $f(\mathbf{x})$ (or $\mathbf{f}(\mathbf{x})$ for an emulator with multiple outputs). In a supervised setup, the performance of the network is quantified by a loss function that measures the difference between the expected outputs, corresponding to the outputs of the training set $\{\mathbf{y_i}\}$, and the outputs computed by the emulator $\{\mathbf{f}(\mathbf{x_i})\}$. Training the emulator implies minimizing the loss function with respect to the emulator parameters, i.e., the weights and biases of all the neurons in the network. Section~\ref{HM} describes the loss functions that are optimized, which are mean squared error (MSE) functions modified to consider the uncertainty prediction explicitly\footnote{Note that in order to evaluate the UQ loss function, the output layer of the model has to be modified as well to include the additional outputs that encode the uncertainty estimation.}.	
	
	\subsection{Heteroscedastic Approach}\label{HM}
	
	A heteroscedastic uncertainty estimate assigns a different uncertainty to each sample. Specifically, the heteroscedastic formulation assumes that the prediction can be modeled as a normal random variable with domain-dependent parameters. For a regression emulator of only one output, the ML emulator learns two outputs: the regular regression prediction $f(\mathbf{x_i})$, which corresponds to the mean, and the uncertainty represented by an additional output $\sigma(\mathbf{x_i})^2$, which corresponds to the variance of the normal distribution. 
	The emulator is trained by minimizing the heteroscedastic loss over the entire dataset,
	\begin{equation}
	\mathcal{L}(y, f, \sigma) = \frac{1}{N} \sum_{i=1}^N  \frac{1}{2 \sigma(\mathbf{x_i})^2} \left \Vert y_i - f(\mathbf{x_i})\right \Vert^2 + \frac{1}{2} \log \sigma(\mathbf{x_i})^2\, . \label{eq:het1D}
	\end{equation}
	This loss function, patterned after the negative log-likelihood of an univariate normal distribution, is composed of two terms: a first term that is a weighted mean squared error (MSE), where large error predictions are compensated by large variances, and a second term that penalizes those large variances. Note that the loss function uses the fact that the expected output $y_i$ for input $\mathbf{x_i}$ is known, allowing for an error term, while the uncertainty $\sigma_i$ is learned indirectly via regularization, i.e. the penalization term that prevents the estimation of variances that are non commensurate with the mean predictions. The balance between the first and second loss terms allows determining an `optimal' uncertainty prediction. Hence, while the output prediction can exploit the available data explicitly, the uncertainty prediction exploits it implicitly. In some cases, this implicit estimation may correspond to a much harder task, and underlying assumptions, like model smoothness, play a greater role in determining the kind of functions that can be induced as uncertainty estimators under this formulation.
	
	\subsubsection{Multivariate Approach} \label{MM}
	When the prediction includes multiple outputs, a one-dimensional normal random variable approach may not be enough to capture the influence of each in the uncertainty prediction of the others. We now show that a multivariate approach that captures the dependence of the two main variables on each other provides a better estimate of the uncertainty. To this end, it is necessary to adapt the formulation of the heteroscedastic loss to consider the probability density function (PDF) of a multivariate normal distribution. The multivariate normal distribution PDF can be written as
	\begin{equation}
	\mathcal{N}(\boldsymbol{\mu}, \Sigma) = \frac{\mathrm{det}(\Sigma)^{-1/2}}{2\pi^{k/2}} \; \exp \left ( - \frac{1}{2} (\mathbf{z} - \boldsymbol{\mu})^T \Sigma^{-1} (\mathbf{z} - \boldsymbol{\mu}) \right)\, , \label{eq:pdf}
	\end{equation}
	where $\boldsymbol{\mu} \in \mathbb{R}^k$ stands for the mean, $\Sigma \in \mathbb{R}^{k \times k}$ represents the positive covariance matrix, and $k$ is the space dimension. The multivariate heteroscedastic loss is expressed then as the negative log-likelihood (NLL) of the multivariate PDF with parameters promoted to functions of the input,
	\begin{eqnarray}
	\mathcal{L}(\mathbf{y}, \mathbf{f}, \Sigma) &=& \nonumber \\ \omit\span\omit\span \frac{1}{N} \sum_{i=1}^N \left ( (\mathbf{y_i} - \mathbf{f}(\mathbf{x_i}))^T \Sigma(\mathbf{x_i})^{-1}(\mathbf{y_i} - \mathbf{f}(\mathbf{x_i})) \right. \nonumber\\
	\omit\span\omit\span \left. {}+ \log \mathrm{det}(\Sigma(\mathbf{x_i}))  \right)\, ,
	\label{eq:hetmultiD}
	\end{eqnarray}
	where constants $(2 \pi)^{-k/2}$ and $1/2$, have been omitted because they do not change the optimum. 
	
	Correspondingly, this approach has to be designed to predict a vector $\mathbf{f}(\mathbf{x_i}) \in \mathbb{R}^k$, with the $k$ outputs of the regular emulator and a matrix $\Sigma(\mathbf{x_i}) \in \mathbb{R}^{k \times k}$ representing the covariance matrix for the UQ estimation.

	The critical component of the multivariate heteroscedastic approach is being able to guarantee that the covariance matrix, $\Sigma$, is symmetric and positive definite. An efficient strategy to achieve this is to formulate the learning task such that a matrix $A^T A$, which is positive by definition, is learned instead. This is a general strategy that can be applied to produce uncertainty estimates for ML emulators with two or more outputs.

	\subsubsection{The Two-Outputs Case}
	In this section, we show the specific structure of the task for a two-output emulator, since this is the case of interest for this work.
	
	The heteroscedastic UQ for an emulator with two outputs has the following structure:
	\begin{itemize}
		\item Base outputs: two to predict the mapping $\mathbf{f}(\mathbf{x_i}) \in \mathbb{R}^2$,
		\item Additional outputs: four to predict the components of matrix $A(\mathbf{x_i}) \allowbreak \in \mathbb{R}^{2 \times 2}$. For simplicity, the explicit $\mathbf{x_i}$ dependence of $A$ is (mostly) omitted in the following description.
	\end{itemize}
	The $2 \times 2$ matrix $A$ can be represented in terms of scalar components $a, b, c, d$, as
	\begin{equation*}
	A = \left ( \begin{array}{cc} a & b \\ c & d \end{array} \right )\, ,
	\end{equation*}
	which in turn yields the covariance matrix
	\begin{eqnarray}
	\Sigma & = & A^T A = \left ( \begin{array}{cc} a & b \\ c & d \end{array} \right )^T \left ( \begin{array}{cc} a & b \\ c & d \end{array} \right ) \nonumber \\
	& = & \left ( \begin{array}{cc} a^2 + c^2 & ab + cd \\ ab + cd & b^2 + d^2 \end{array} \right ) 
	\triangleq \left( \begin{array}{cc} \sigma_{11}^2 & \sigma_{12} \\ \sigma_{12} & \sigma_{22}^2 \end{array} \right )\, . \nonumber\\ \label{eq:cov}
	\end{eqnarray}
	Since any simultaneous rotation of the vectors given by the columns of matrix $A$ by the same angle leaves the covariance matrix unchanged, we choose rotation angle $\theta = \arctan (c-b)/(a+d)$, to force $b = c$. Thus, the UQ approach is built to have only five outputs: two for $\mathbf{f}(\mathbf{x_i})$ and three for the distinct components of matrix $\Sigma(\mathbf{x_i})$: $\sigma_{11}(\mathbf{x_i}), \sigma_{12}(\mathbf{x_i}), \sigma_{22}(\mathbf{x_i})$. 
	
	Further simplifications can be achieved by applying the following observations. For a positive definite covariance matrix $\Sigma \in \mathbb{R}^{2 \times 2}$, the inverse can be computed analytically as
	\begin{equation}
	\Sigma^{-1} = \frac{1}{\mathrm{det}(\Sigma)} \; \left( \begin{array}{cc} \sigma_{22}^2 & -\sigma_{12} \\ -\sigma_{12} & \sigma_{11}^2 \end{array} \right )\, ,
	\label{eq:invcov}
	\end{equation}
	with $\mathrm{det}(\Sigma) = \sigma_{11}^2 \sigma_{22}^2 - \sigma_{12}^2 \neq 0$. Defining: $\mathbf{e} = (e_1, e_2)^T = \mathbf{y} - \mathbf{f}(\mathbf{x})$, i.e., the difference between the ground truth $\mathbf{y}$ and the mean prediction $\mathbf{f}(\mathbf{x})$, allows to write
	\begin{eqnarray}
	\mathrm{NLL}  = \frac{1}{\mathrm{det}(\Sigma)} \left( \begin{array}{c} e_1 \\ e_2 \end{array} \right )^T \left ( \begin{array}{cc} \sigma_{22}^2  -\sigma_{12} \\ -\sigma_{12} \sigma_{11}^2 \end{array} \right ) \left( \begin{array}{c} e_1 \\ e_2 \end{array} \right ) \nonumber\\
	\qquad {} + \log{\mathrm{det}(\Sigma)} \nonumber \\
	= \frac{1}{\mathrm{det}(\Sigma)} ( \sigma_{22}^2 \; e_1^2 - 2 \; \sigma_{12} \; e_1 \; e_2 + \sigma_{11}^2 \; e_2^2 ) + \log{\mathrm{det}(\Sigma)}\, .\nonumber\\
	\label{eq:nll2D}
	\end{eqnarray}
	Hence, the loss function for a heteroscedastic approach with two outputs corresponds to
	\begin{eqnarray}
	\mathcal{L}(\mathbf{y}, \mathbf{f}, \Sigma) = \frac{1}{N} \sum_{i=1}^N  \left ( \frac{1}{\mathrm{det}(\Sigma(\mathbf{x_i}))} ( \sigma_{22}(\mathbf{x_i})^2 (y_1 - f_1(\mathbf{x_i}))^2 \right . \nonumber \\
	\quad - \; 2 \; \sigma_{12}(\mathbf{x_i}) (y_1 - f_1(\mathbf{x_i})) (y_2 - f_2(\mathbf{x_i})) \nonumber \\
	\left . \phantom{\frac{1}{1}} + \sigma_{11}(\mathbf{x_i})^2 \; (y_2 - f_2(\mathbf{x_i}))^2 ) + \log{\mathrm{det}(\Sigma(\mathbf{x_i}))} \right )\, .\nonumber\\ \label{eq:het2D}
	\end{eqnarray}

\section{Application to the Flyer Plate Problem}

Remember that the focus of this work is to train a ML emulator to estimate the uncertainty in the one--time-step response prediction of two quantities of interest for the flyer plate problem described in Section~\ref{sec:problem}. For this purpose, a feed-forward neural network was constructed, and its performance was compared against different experimental setups.

The input of the emulator constructed in this work corresponds to a sequence of $d$ pairs of consecutive time steps of the length of the longest crack, $L_{\mathrm{long}}$, and the maximum tensile stress, $S_{yy}$. The emulator predicts the next step in the evolution of these two quantities as well as the associated uncertainties. In other words, given the input represented by the $2d$ sequence of time steps $(1, \dots, d)$ of the length of the longest crack and the maximum tensile stress,
\begin{equation}
	\left [ \left ( L_{\mathrm{long}}^{(1)}, S_{yy}^{(1)} \right ), \left ( L_{\mathrm{long}}^{(2)}, S_{yy}^{(2)} \right ), \dots, \left ( L_{\mathrm{long}}^{(d)}, S_{yy}^{(d)} \right ) \right ]\, ,
\end{equation}
the emulator predicts: $\left ( L_{\mathrm{long}}^{(d+1)}, S_{yy}^{(d+1)} \right )$, as well as the covariance matrix $\Sigma^{(d+1)}$ for these two quantities, which is composed of variances: ${\sigma^2_{L_{\mathrm{long}}}}^{(d+1)}$ and ${\sigma^2_{S_{yy}}}^{(d+1)}$, and covariance: ${\sigma_{L_{\mathrm{long}},S_{yy}}}^{(d+1)}$. The window is advanced by one time step, such that the input corresponds now to quantities between $t=2$ and $d+1$ and the next time step $d+2$ is predicted. The process of advancing the input window is repeated until the whole evolution sequence is predicted.

  \subsection{Simulation Data}
\begin{figure}[!htbp] 
		\centering
		\subfigure[Evolution of length of the longest crack $L_{\mathrm{long}}$. \label{fig:qoi1}]
		{\includegraphics[width=0.85\columnwidth]{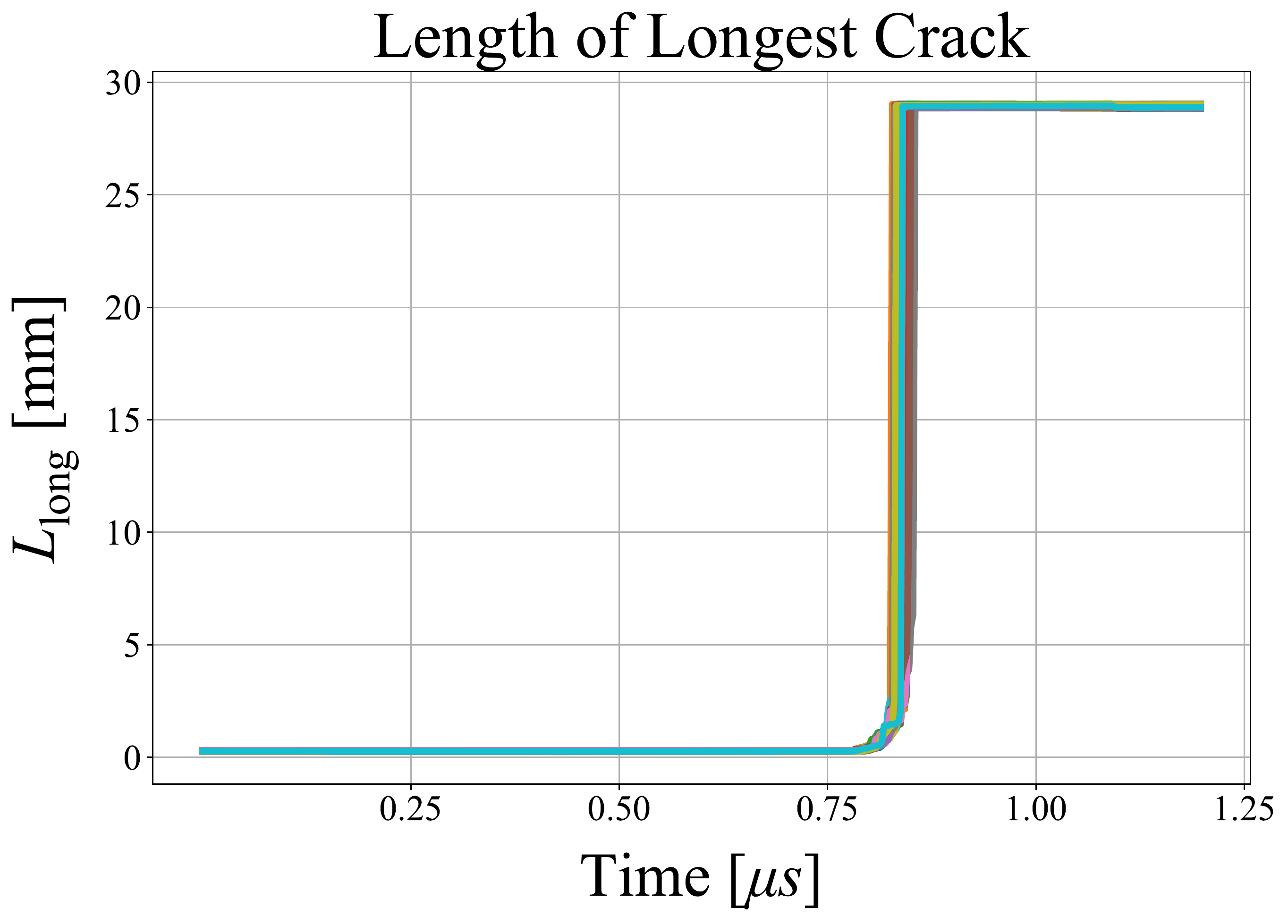} } \\
		\subfigure[Evolution of maximum tensile stress $S_{yy}$. \label{fig:qoi2}]
		{\includegraphics[width=0.85\columnwidth]{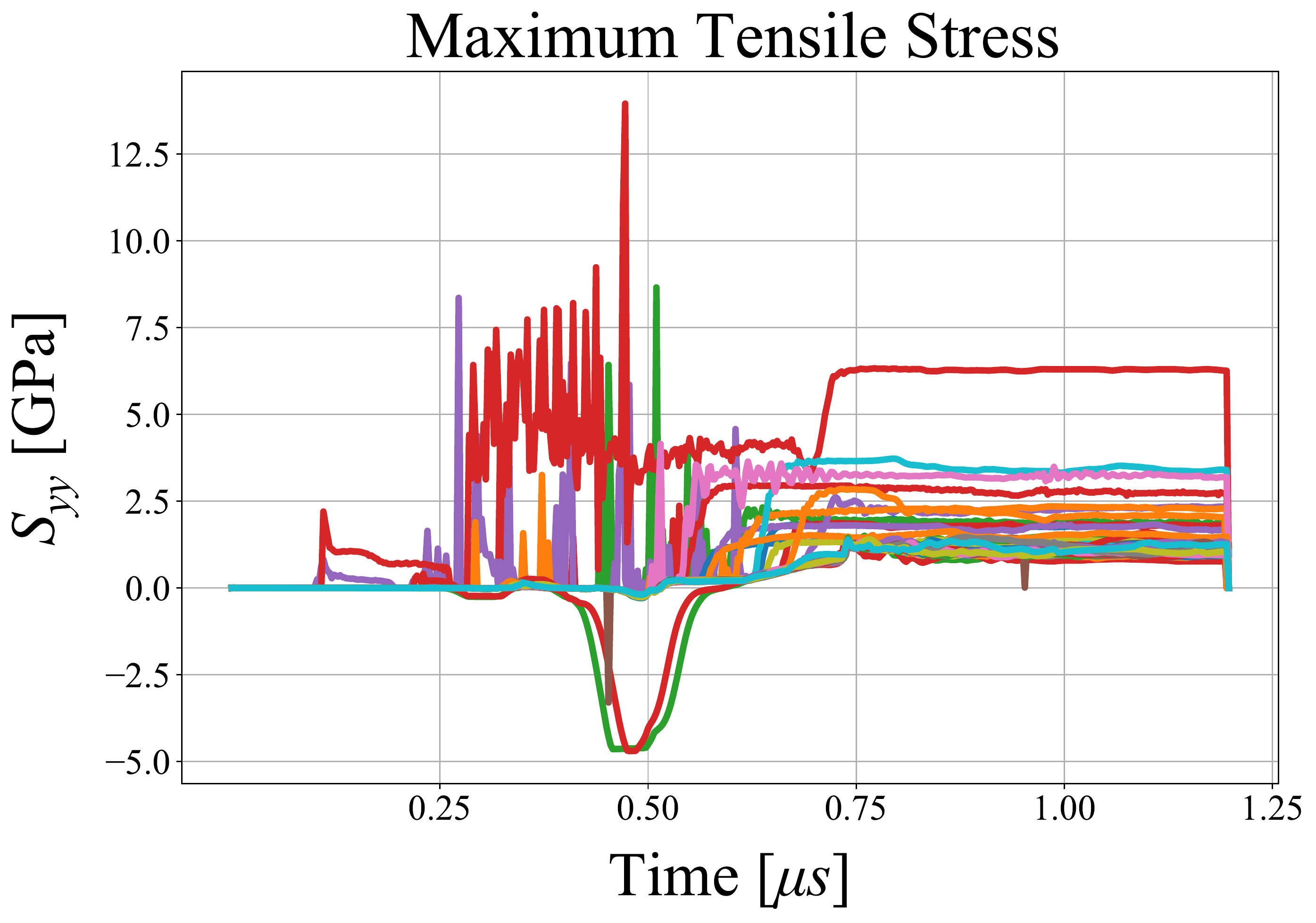} }
		\caption{The 100 different HOSS simulations used for train and test the ML emulator. \label{fig:qois}}
	\end{figure}

	Similarly to~\cite{fernandez2021accelerating}, 100 HOSS flyer plate simulations described in Sec.~\ref{sec:HOSS} were used. Each simulation includes time series of 480 time steps for different quantities of interest. From each of the 100 simulations the outputs of interest, $L_{\mathrm{long}}$ and $S_{yy}$, are extracted. Figure~\ref{fig:qoi1} and Figure~\ref{fig:qoi2} show $L_{\mathrm{long}}$ and $S_{yy}$, respectively as a function of time. As Figure~\ref{fig:qois} shows, the length of the longest crack as a function of time is less sensitive to the considered variations in the inputs than the maximum tensile stress as a function of time.
	
	Since the task is one--time-step prediction, each time series is split into subsequences containing $d + 1$ time steps each, where the first $d$ constitutes the input pattern $\mathbf{x}$ and the last the output $\mathbf{y}$. Remember that because we are interested in the $L_{\mathrm{long}}$ and $S_{yy}$ interaction, the input subsequences are paired as in Eq.~\eqref{eq:x}, while the output consists of pairs as in Eq.~\eqref{eq:y}, with dimensionalities $\mathbf{x} \in \mathbb{R}^{2d}$ and $\mathbf{y} \in \mathbb{R}^2$, respectively. To guarantee that subsequences of the same simulation are not mixed during training and testing, they are first split randomly into 80\% simulations for training and 20\% simulations for testing. Afterward, they are further split into subsequences to build the corresponding sets, keeping subsequences of the training simulations in the training set and subsequences of the testing simulations in the testing set. Different embedding dimensions $d$ are used. The sizes of the resulting sets are summarized in Table~\ref{tab:data}. Note that $d$ input components are needed to predict the $d+1$ component, meaning that some of the left-most values in the simulation do not have enough previous elements to build the input pattern. Consequently, the resulting number of usable subsequences slightly decreases when $d$ increases. 
	
	\begin{table}[!h]
		\centering
		\begin{footnotesize}
			\renewcommand{\arraystretch}{1}	
			\setlength{\tabcolsep}{3pt}
			\begin{tabular}{ |c|c|c|c|}
				\hline
				\multicolumn{1}{|c|}{ {\bf Embedding Dimension} } & \multicolumn{1}{|c|}{ {\bf Subsequences} } & \multicolumn{1}{|c|}{ {\bf Training Set} } &
				\multicolumn{1}{|c|}{ {\bf Testing Set} } \\
				\hline
				\hline
				$d=$ {\bf 10} & 469 & 37,520 & 9,380
				\\
				\hline
				$d=$ {\bf 20} & 459 & 36,720 & 9,180
				\\
				\hline
				$d=$ {\bf 30} & 449 & 35,920 & 8,980
				\\
				\hline
			\end{tabular}
		\end{footnotesize}
		\caption{\small Data sets for one-time-step prediction.}
		\label{tab:data}
	\end{table}
	
	Although a set of 100 HOSS flyer plate simulations may be perceived as a relatively small dataset, we remark that its decomposition into one-time-step predictions generates a considerable amount of training data as shown in Table~\ref{tab:data}. Previous results obtained by our group evidence that heteroscedastic UQ models have good performance even in the low data limit~\cite{garcia-2021-low} but a full analysis of the data requirements is beyond the scope of this work.

	\subsection{Model Architecture and Training}
	A feed-forward neural network with two hidden layers of 200 neurons each and output layer of five neurons was constructed. A different emulator was trained for each of the different embedding dimensions. The corresponding networks have 45,405; 49,405; and 53,405 parameters, respectively. Each emulator is trained for 50 epochs with batch size of 20 using an Adam optimizer. The neural network emulators were built and trained with the Python package Keras~\cite{chollet2015keras}. Training one of the machine learning emulator models with the 2D heteroscedastic approach takes about 425s (7.1 minutes) and evaluating the testing set about 0.5s in a MacBook Pro (2.4 GHz 8-Core Intel Core i9) using CPU only. The training process is repeated 20 times, using different training-testing partitions. 
	
	Note that we did not attempt to optimize the architecture of the emulator, we focused instead on assessing the efficacy of the multivariate heteroscedastic formulation for models that have reasonable performance in the basic prediction task (which is evidenced by the relatively high $R^2$ values obtained as described in following sections) and that exhibit stable convergence for multiple random initializations.
	
	\subsection{Model Performance}
	
	 To quantify the emulator performance, the coefficient of determination ($R^2$),
	\begin{equation}
	R^2(y, \hat{y}) = 1 - \frac{\sum_{i=1}^n \left ( y_i - \hat{y}_i \right )^2}{\sum_{i=1}^n \left ( y_i - \bar{y} \right )^2} \; ,
	\quad
	\bar{y} =  \frac{1}{n} \sum_{i=1}^n y_i \, ,
	\end{equation}
	was used, as computed by the Scikit-learn Python package \cite{scikit-learn} and reported in Table~\ref{tab:r2} as the average and standard deviation over the 20 repetitions evaluated in the testing set (i.e. the set held out during training). As the table shows, the performance in terms of $R^2$ is good, specially for $L_{\mathrm{long}}$, and the predictions with embedding dimension $d=20$ is slightly better than the other two cases. 
	
	These performance results and the low training/testing times required, together with the performance achieved by other neural network-based models such as~\cite{fernandez2021accelerating}, demonstrate that the ML emulator approach provides good accuracy with a significant speed-up gain after the training process is complete. Ultimately the goal of this work is to accelerate uncertainty quantification workflows where the number of model runs greatly exceeds the number of model runs used in training. In such cases, this approach will accelerate the workflow even when the cost of the training data is included.
	
	\begin{table}[!h]
		\centering
		\begin{footnotesize}
			\renewcommand{\arraystretch}{1}
			\setlength{\tabcolsep}{3pt}
			\begin{tabular}{ |c|c|c|}
				\cline{2-3}
				\multicolumn{1}{c}{ } & \multicolumn{2}{|c|}{ $\boldsymbol{R^2}$ }  \\
				\hline
				\multicolumn{1}{|c|}{ {\bf Embedding Dimension}  } & \multicolumn{1}{|c|}{ $\boldsymbol{L_{\mathrm{long}}}$ }  & \multicolumn{1}{|c|}{ $\boldsymbol{S_{yy}}$ }  \\
				\hline
				$d=$ {\bf 10} & 0.97 $\pm$ 0.03 & 0.87 $\pm$ 0.22
				\\
				\hline
				$d=$ {\bf 20} & 0.97 $\pm$ 0.02 & 0.89 $\pm$ 0.09
				\\
				\hline
				$d=$ {\bf 30} & 0.95 $\pm$ 0.05 & 0.86 $\pm$ 0.10
				\\
				\hline
			\end{tabular}
		\end{footnotesize}
		\caption{\small $R^2$: Mean $\pm$ standard deviation for two-output models.}
		\label{tab:r2}
	\end{table}

	\subsubsection*{Estimating the Emulator Coverage}
	
	The $R^2$ score only takes into account the predicted values $\mathbf{f}(\mathbf{x})$, so another metric is required to evaluate the predicted uncertainty. For this purpose, we estimate the emulator coverage, which evaluates how many times the prediction falls inside the confidence interval corresponding to a specified ventile\footnote{The $x^{\rm th}$ ventile bounds the region where $x$-twentieth of the data are predicted to lie~\cite{wilcox2016qtl}.}. The closer the fraction of points inside the interval and the ventile are, the tighter the predicted uncertainties.
	
	To estimate the emulator coverage, it is necessary to calculate the expected fraction inside specified contour levels of the multivariate normal distribution learned by the emulator. Contours of the multivariate normal distribution are the set of values where the argument of the exponential in the PDF is the same. As described in Appendix~\ref{sec:app}, each contour corresponds to an ellipse for data in $\mathbb{R}^2$, which can be expressed as 
\begin{equation*}	
	\gamma = (\mathbf{z} - \boldsymbol{\mu})^T \Sigma^{-1} (\mathbf{z} - \boldsymbol{\mu}) \:, \quad \gamma = - 2 \ln \left(1- \alpha \right) \:,
\end{equation*}
with $\gamma$ corresponding to the level of the contour and $\alpha$ to the confidence level. 
	
\subsection{Results}
     	To understand the need of the multivariate approach, we first demonstrate the one-variable heteroscedastic approach that independently estimates the standard deviation for the quantities of interest: length of the longest crack, $L_{\mathrm{long}}$, and maximum tensile stress, $S_{yy}$, in the current setting. Note that the architecture of this one-variable model is similar to the two-output case, however the output layer includes four outputs for the two means and the two variances of the quantities of interest and the loss function used for training is a sum of Eq.~\ref{eq:het1D} applied to each of them individually. Table~\ref{tab:r2_indep} reports the corresponding $R^2$ for 20 different model realizations. This table shows similar levels of $R^2$ for both independent predictions, but these one-output models cannot be used to quantify the interactions between the variables of interest.
	
\begin{figure}[tbp] 
		\centering
		\subfigure[$d=10$.\label{fig:fraccov1_1D}]
		{\includegraphics[width=0.75\columnwidth]{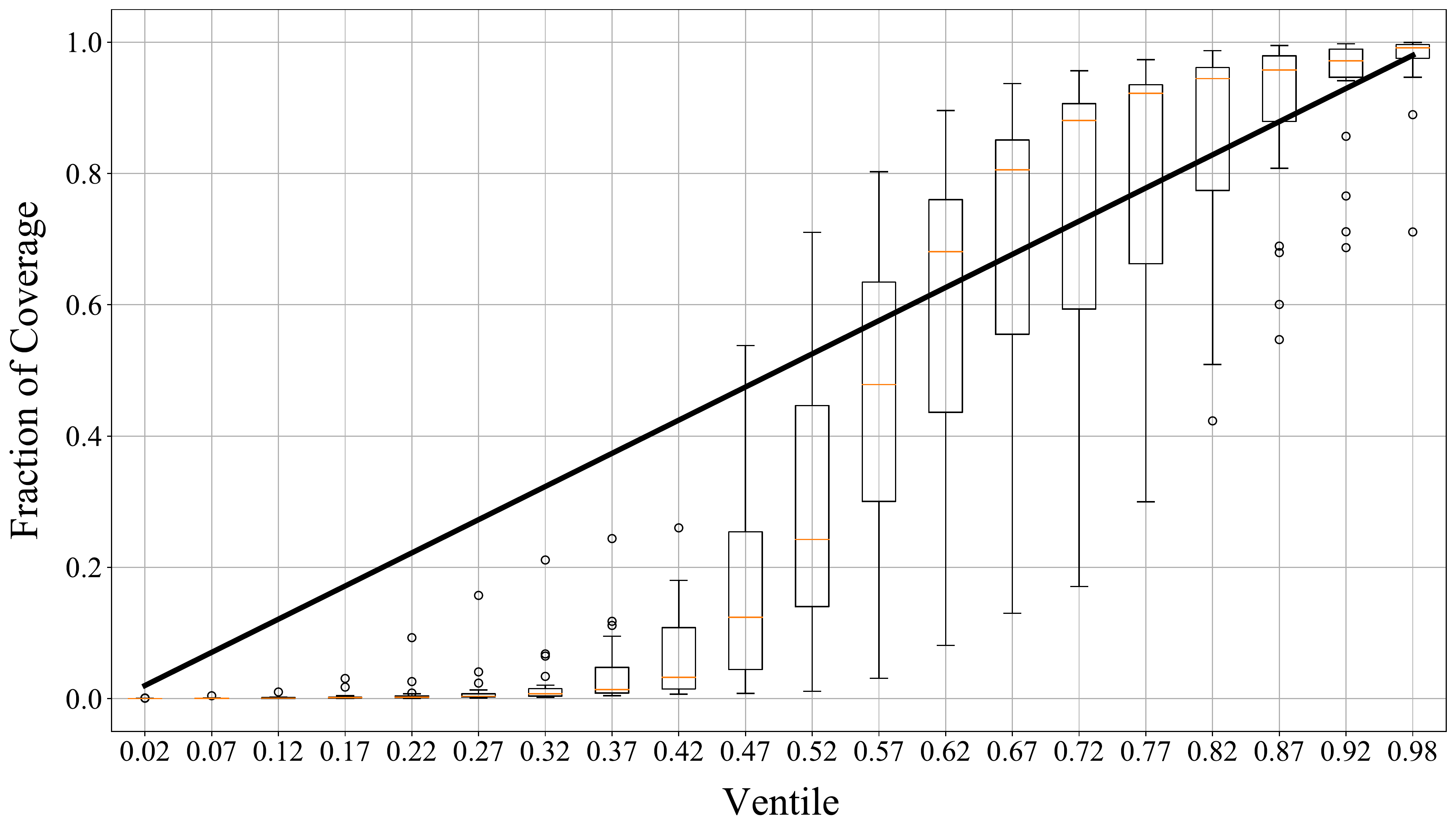} } \\
		\subfigure[$d=20$. \label{fig:fraccov2_1D}]
		{\includegraphics[width=0.75\columnwidth]{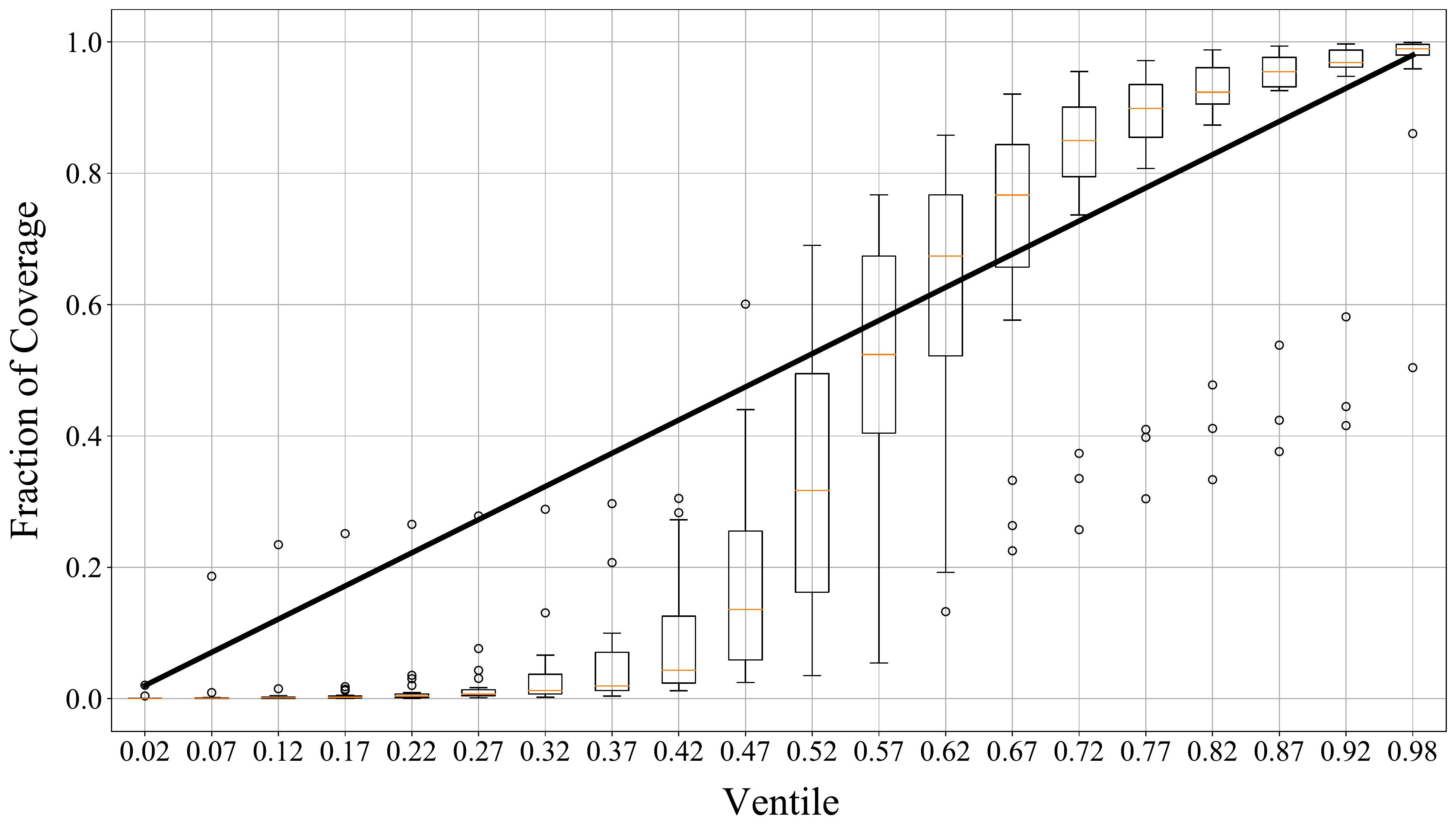} } \\
		\subfigure[$d=30$. \label{fig:fraccov3_1D}]
		{\includegraphics[width=0.75\columnwidth]{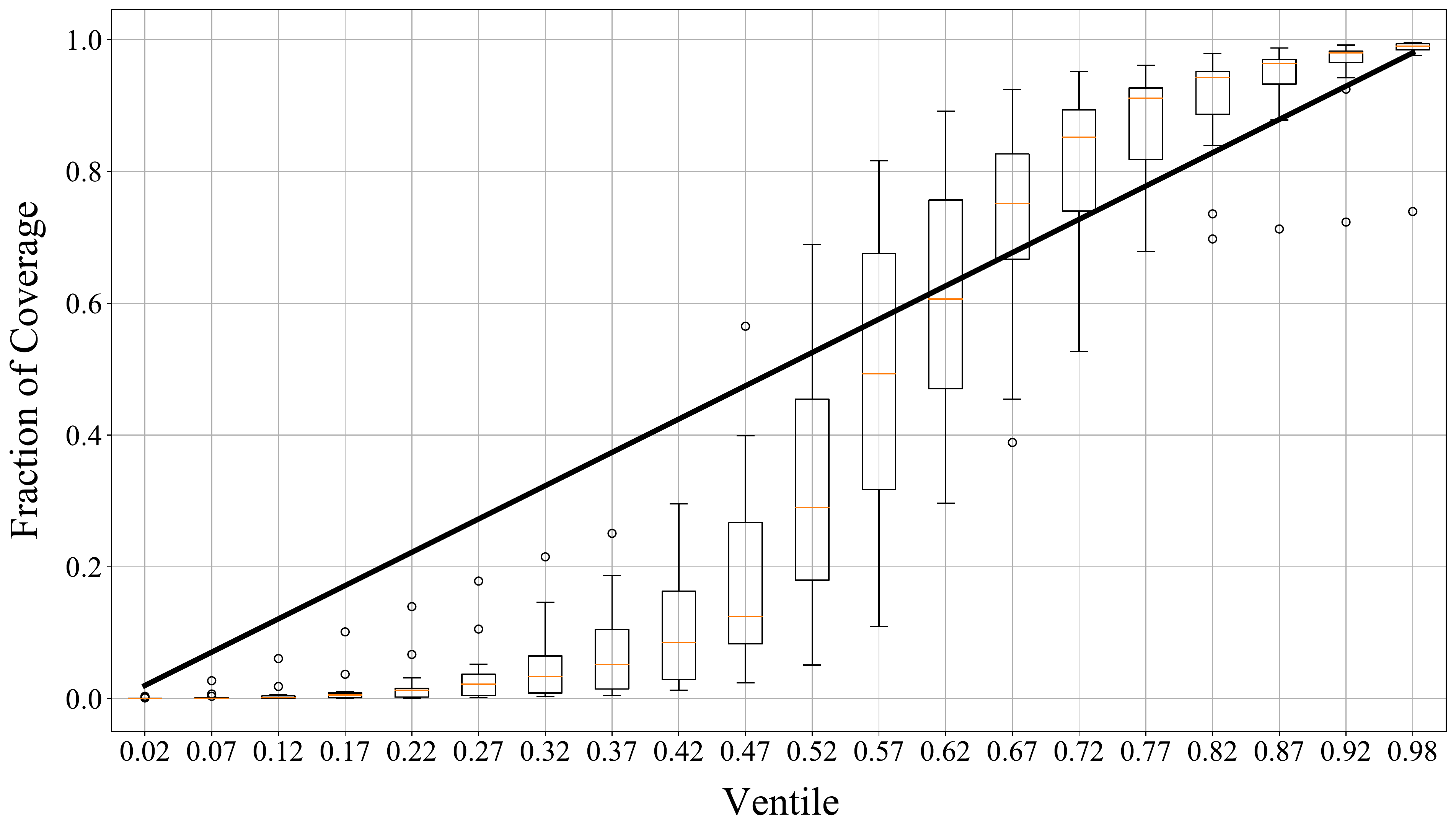} }
		\caption{Box plots of fraction of coverage for $d=10, 20, 30$ over 20 repetitions of the one-variable heteroscedastic approach. \label{fig:fraccov_1D}}
	\end{figure}

	The standard deviations can be used to compute the expected distribution ventiles, which in turn can be compared to the fraction of ground truth samples that effectively fall in the given ventile. Since, in this case, the analysis is carried out independently, both independent conditions are checked simultaneously, {\it i.e.,} a sample is said to fall in a given ventile $V$ if and only if ${L_{\mathrm{long}}}_j$ falls in ventile $V_{L_{\mathrm{long}}}$ computed for the distribution of $L_{\mathrm{long}}$ and ${S_{yy}}_j$ falls in ventile $V_{{S_{yy}}}$ computed for the distribution of $S_{yy}$. Figure~\ref{fig:fraccov_1D} shows the fraction of ground truth samples effectively falling in a given ventile. This is represented by box plots, based on 20 repetitions, for the emulator coverage computed on the testing sets as a function of the specified ventile. The meaning of each box plot is as follows: the box is plotted between the first and third quartiles, the orange line inside the box is the median, the difference between the third and the first quartile is the interquartile range (IQR). The whiskers extend between a distance of 1.5 times the IQR below the lower quartile and a distance of 1.5 times the IQR above the upper quartile. Other observed points outside the whiskers are plotted as outliers. Additionally, the ideal relationship between fraction of coverage and ventile is plotted as a continuous black line. It can be seen that in all cases the fraction of coverage exhibits an S-shape with slight over-prediction in larger ventiles and significant under-prediction in smaller ventiles.

\begin{table}[!htbp]
		\centering
		\begin{footnotesize}
			\renewcommand{\arraystretch}{1}
			\setlength{\tabcolsep}{3pt}
			\begin{tabular}{ |c|c|c|}
				\cline{2-3}
				\multicolumn{1}{c}{ } & \multicolumn{2}{|c|}{ $\boldsymbol{R^2}$ }  \\
				\hline
				\multicolumn{1}{|c|}{ {\bf Embedding Dimension}  } & \multicolumn{1}{|c|}{ $\boldsymbol{L_{\mathrm{long}}}$ }  & \multicolumn{1}{|c|}{ $\boldsymbol{S_{yy}}$ }  \\
				\hline
				$d=$ {\bf 10} & 0.96 $\pm$ 0.04 & 0.98 $\pm$ 0.004
				\\
				\hline
				$d=$ {\bf 20} & 0.93 $\pm$ 0.11 & 0.96 $\pm$ 0.049
				\\
				\hline
				$d=$ {\bf 30} & 0.93 $\pm$ 0.06 & 0.98 $\pm$ 0.004
				\\
				\hline
			\end{tabular}
		\end{footnotesize}
		\caption{\small $R^2$: Mean $\pm$ standard deviation for one-output models.}
		\label{tab:r2_indep}
	\end{table}

\begin{figure}[!htbp] 
		\centering
		\includegraphics[width=0.95\columnwidth]{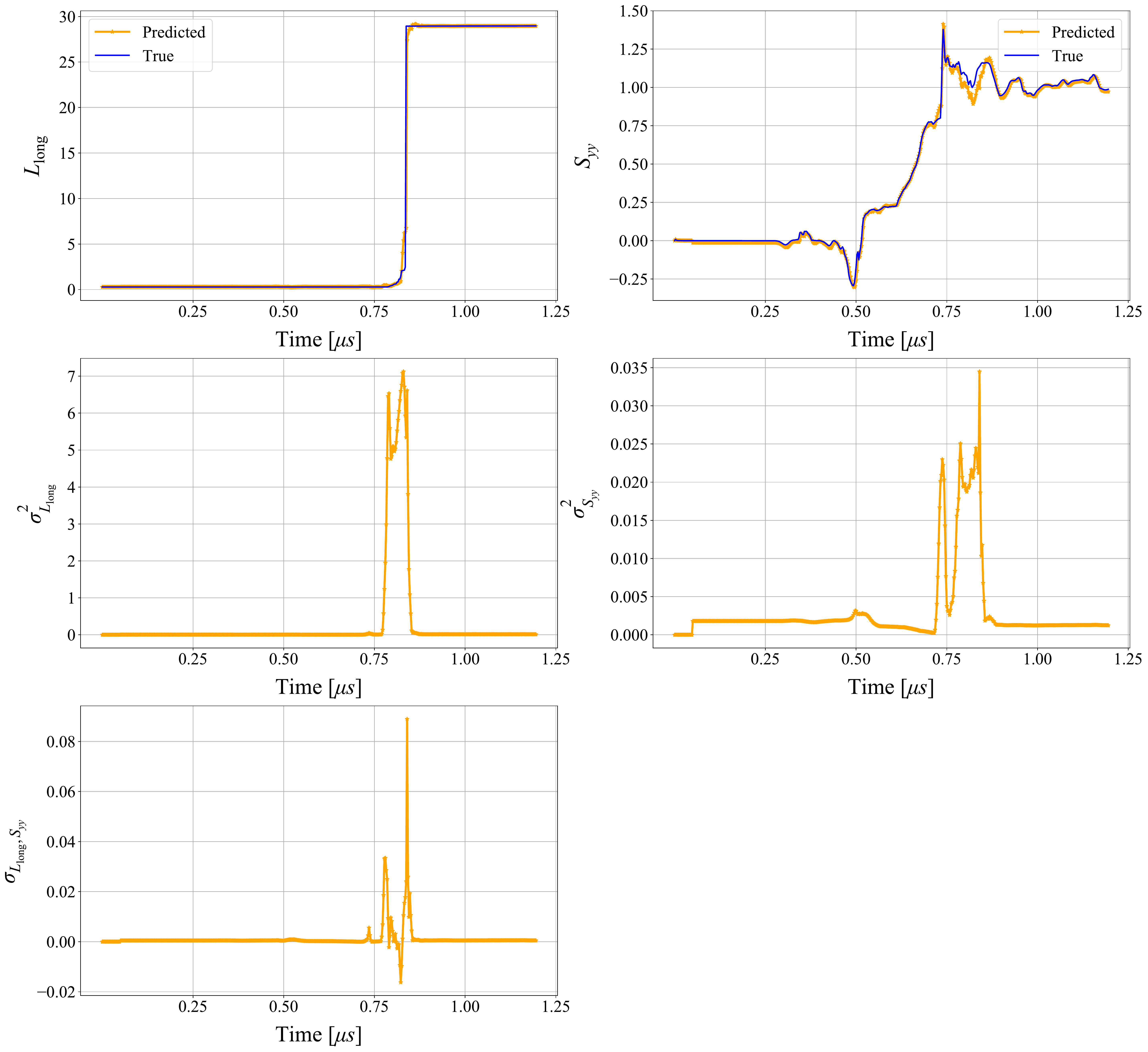}
		\caption{Mean and covariances predicted for one of the testing series for one of the trained models for $d=20$. \label{fig:pred_2D}}
\end{figure}

	The underprediction in the lower ventiles is a significant concern, and could result from a correlation between the two quantities. And, indeed, a quantitative analysis of the covariance matrix resulting from the multivariate analysis and plotted in Figure~\ref{fig:pred_2D} shows that the larger scale is observed for the variance of the length of the longest crack $\sigma^2_{L_{\mathrm{long}}}$, but this is manifest only in a window surrounding the sharp transition in the crack length. On the other hand, the scales of the variance for the maximum tensile stress $\sigma^2_{S_{yy}}$ and the covariance of length of the longest crack and maximum tensile stress $\sigma_{L_{\mathrm{long}},S_{yy}}$ are comparable, although smaller than $\sigma^2_{L_{\mathrm{long}}}$. The covariance $\sigma_{L_{\mathrm{long}},S_{yy}}$ is also apparent in a window surrounding the sharp transition in the crack length. Qualitatively, this behavior seems consistent, in that the uncertainty in the length of the longest crack is concentrated in the transition region, and that the significant interaction between the variables is also concentrated in the same region. 

	\begin{figure}[!htbp] 
		\centering
		\subfigure[$d=10$.\label{fig:fraccov1}]
		{\includegraphics[width=0.75\columnwidth]{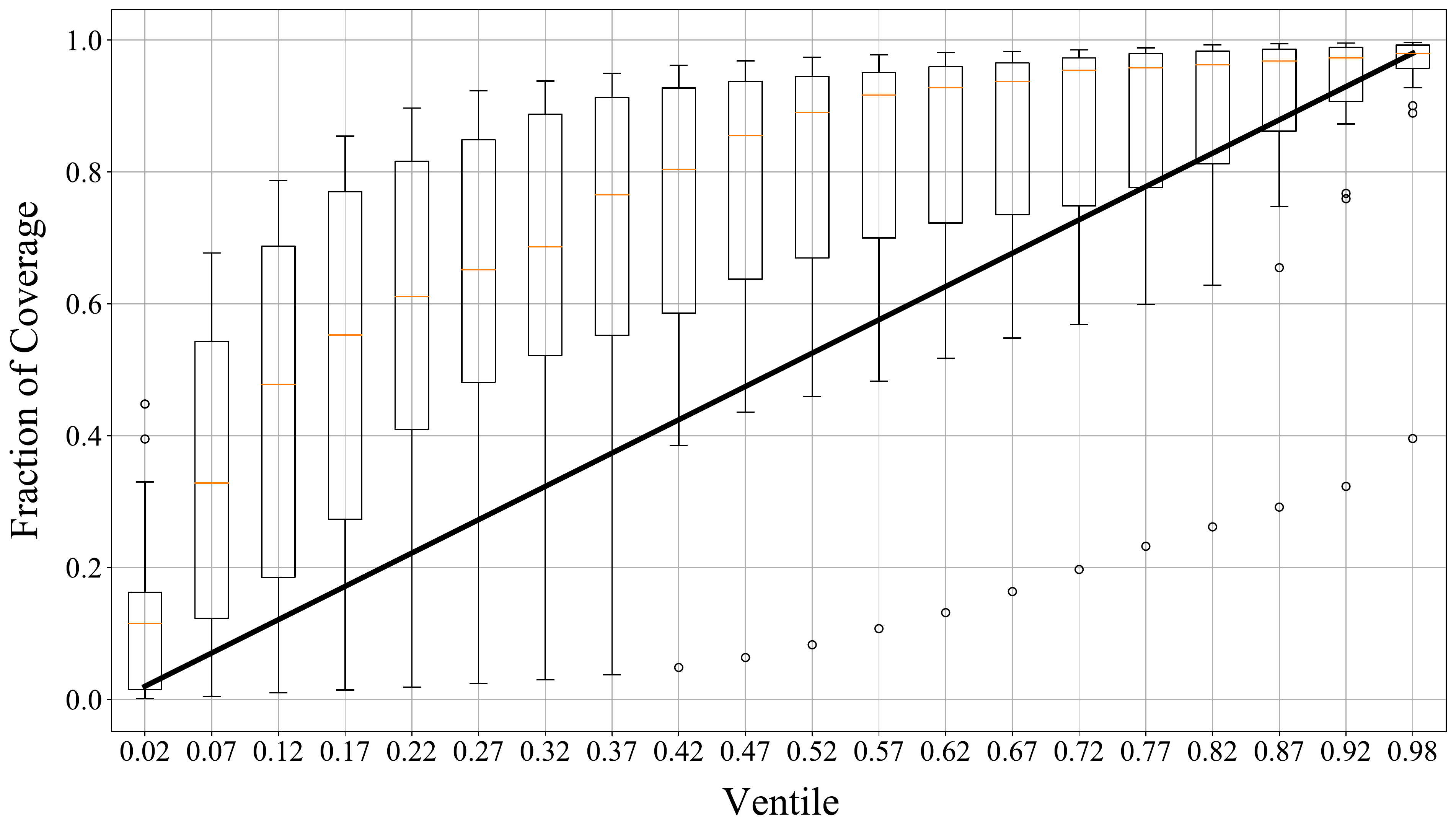} } \\
		\subfigure[$d=20$. \label{fig:fraccov2}]
		{\includegraphics[width=0.75\columnwidth]{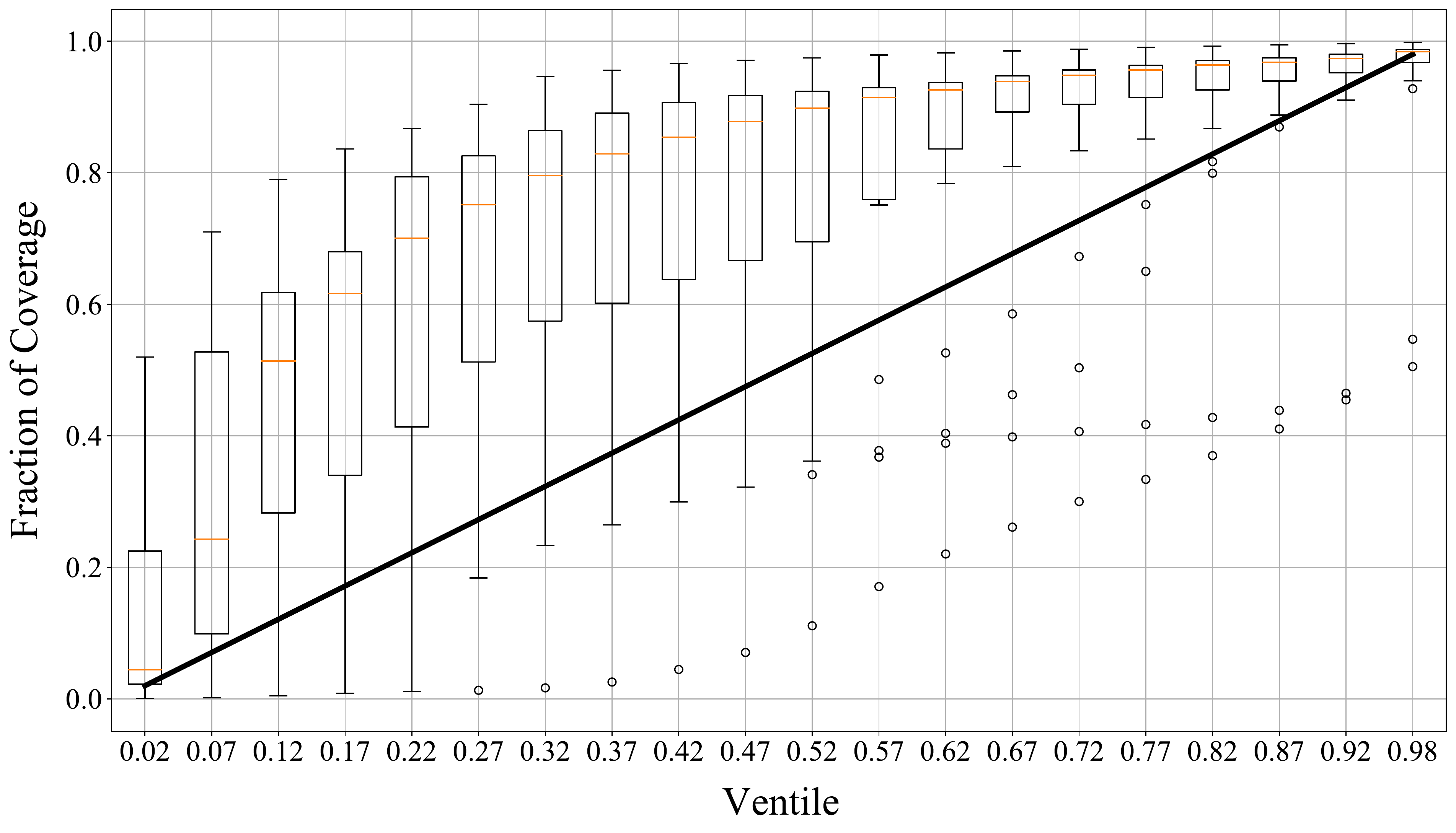} } \\
		\subfigure[$d=30$. \label{fig:fraccov3}]
		{\includegraphics[width=0.75\columnwidth]{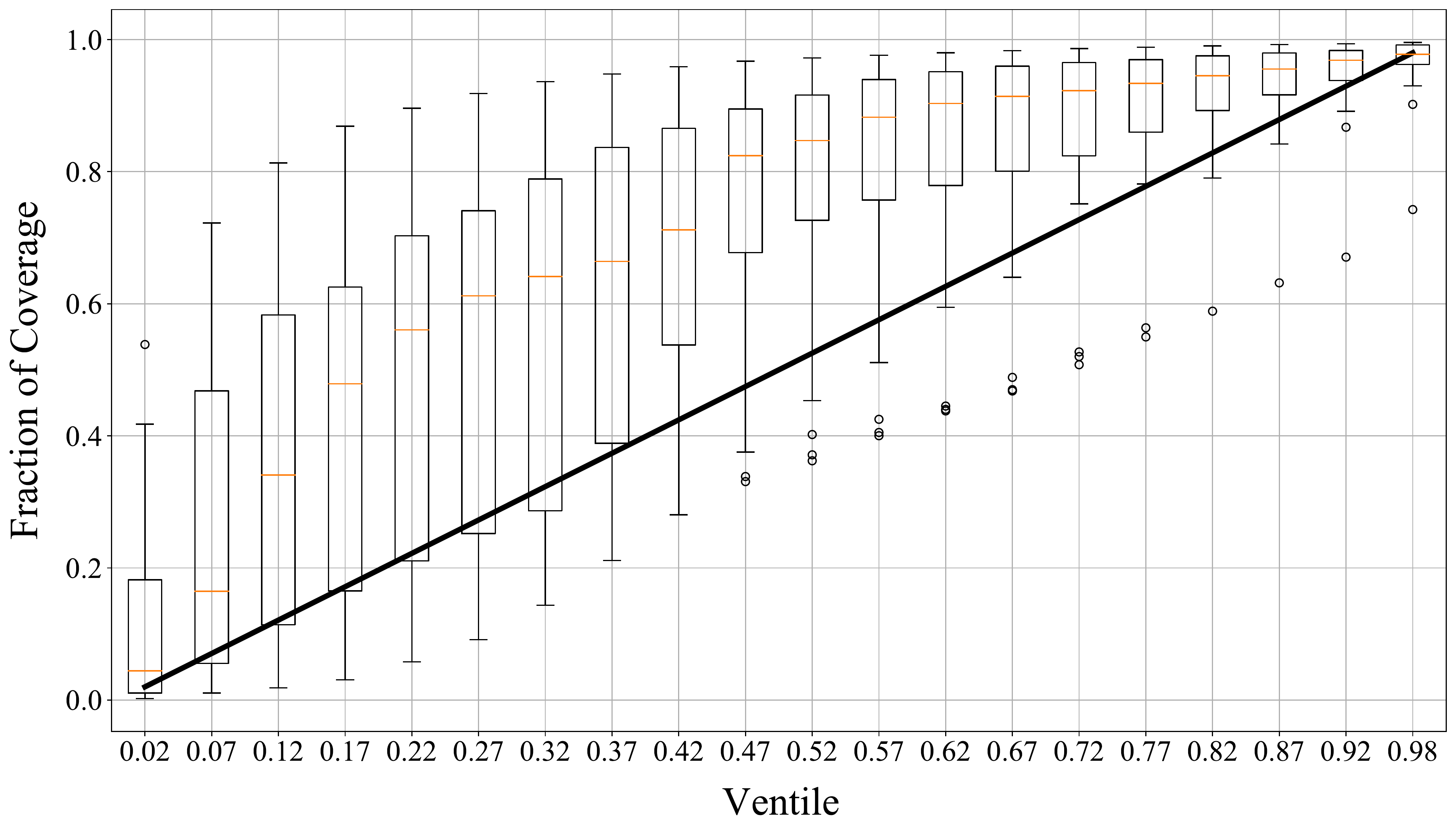} }
		\caption{Box plots of fraction of coverage for $d=10, 20, 30$ over 20 repetitions of the multivariable heteroscedastic approach. \label{fig:fraccov}}
	\end{figure}

	As a result, the uncertainty estimates are qualitatively different from the one produced by the multivariate approach, where the variable correlations are captured.  Figure~\ref{fig:fraccov} displays the evaluation of the emulator coverage in this latter case. It can be seen that all the emulators tend to overestimate the confidence interval, which implies that the uncertainty estimation is conservative, i.e., biased towards the safer side of including more fraction of predictions than what could be inferred from the ventile. Also, the medians observed tend to be closer to the ideal for $d=10$ and $d=30$ than for $d=20$, with the best overall statistics for $d=30$. This illustrates the tension between prediction and uncertainty estimation: the ranking of emulators by $R^2$ may differ from the one obtained by coverage. In practice, we observe that the complexity of the function required for accurate mean prediction may be different to the complexity required for accurate variance prediction. Hence, it may be beneficial to further tune the regularization of the variance in order to avoid an oversmoothed mean prediction or limit the noise in the variance prediction.  
	
	\section{Conclusion}
	Quantifying the uncertainty sources associated with physical models is of high importance for their credibility. When a machine learning emulator is used to speed up the process of predicting crack evolution in high-strain brittle experiments, it is important to evaluate the uncertainty associated with how well the machine learning emulator captures the underlying simulation. To some extent, this mitigates the undesirable ``black box'' nature of machine learning emulators. It does not make the predictions interpretable, but instead gives an indication of the expected accuracy of the prediction.

	The main contribution of this work is to use machine-learning itself to bound the multivariate response of such an emulator using a heteroscedastic approach. The machine learning response is accurate within its predicted errors, while uncertainty predictions conservatively overestimate the coverage for the given confidence levels. Thus, for example, the 95\% confidence interval covers about 97.6\% of the data. This behavior is much more desirable than underpredicting the uncertainty. Underpredicting the uncertainty would make the predictions seem more accurate than they are, which could have serious undesirable consequences in contexts where safety relies on the material response. The underlying cause of this overprediction is probably related to insufficiency of model assumptions ---especially near a failure point--- and points to the need for a nonparametric estimator for the uncertainty. We leave that development to future work.
	
	\printcredits
	
	\section*{Data Availability}
	
	Available from the authors upon request.

	\section*{Acknowledgements}
	The authors are grateful to the anonymous reviewers, whose comments and suggestions helped improve the clarity of the manuscript. 
	
	MGFG and DO acknowledge support from the National Nuclear Security Administration's Advanced Simulation and Computing program. This work has been supported in part by the Joint Design of Advanced Computing Solutions for Cancer (JDACS4C) program established by the U.S. Department of Energy (DOE) and the National Cancer Institute (NCI) of the National Institutes of Health, and was performed under the auspices of the U.S. Department of Energy by \\ Lawrence Livermore National Laboratory under Contract \\ DE-AC52-07NA27344 and Los Alamos National Laboratory under Contract DE-AC5206NA25396. Approved for public release LA-UR-20-30015 and LLNL-JRNL-817876.

	\bibliographystyle{cas-model2-names}
	\bibliography {bibliography}
	
	\appendix
	
	\section{Determination of Contours for Multivariate Normal Distribution} \label{sec:app}
	
	Contours of the multivariate normal distribution are the set of values where the argument of the exponential in the PDF is the same. Therefore, contours correspond to
	\begin{equation*}
	(\mathbf{x} - \boldsymbol{\mu})^T \Sigma^{-1} (\mathbf{x} - \boldsymbol{\mu}) = \gamma \, ,
	\end{equation*}
	where $\gamma > 0$ is a constant value. For data in $\mathbb{R}^2$, each contour corresponds to an ellipse.
	For simplicity it is assumed that $\boldsymbol{\mu} = \mathbf{0}$ and that the covariance matrix has been diagonalized. Therefore, 
	\begin{equation*}
	(\mathbf{x} - \boldsymbol{\mu})^T \Sigma^{-1} (\mathbf{x} - \boldsymbol{\mu}) = \left(\frac{x}{\sigma_x}\right)^2 + \left(\frac{y}{\sigma_y} \right)^2 \, .
	\end{equation*}
	Integrating the PDF of the multivariate inside the ellipse and requesting it to be equal to a specific coverage $\alpha$ yields to
	\begin{eqnarray*}
		4 \int_0^{\sigma_x} \int_0^{\sigma_y \sqrt{\gamma - \left(\frac{x}{\sigma_x}\right)^2}} \frac{1}{2\pi} \; \frac{1}{\sigma_x \sigma_y} \times {}\span\omit\span\\
		\qquad \exp \left [ -\frac{1}{2} \left ( \left(\frac{x}{\sigma_x}\right)^2 + \left(\frac{y}{\sigma_y} \right)^2 \right ) \right ] \; dy \; dx \span\omit\span\\
		\qquad\qquad\qquad&=&  \alpha \, ,
	\end{eqnarray*}
	where the integral is computed over the quarter ellipse in the first quadrant.
	
	Making the change of variables to
	\begin{equation*}
	\begin{array}{c} x = \sigma_x \; z \; \cos \theta \\ y = \sigma_y \; z \; \sin \theta \end{array} \, ,
	\end{equation*}
	and computing the Jacobian of the transformation
	\begin{equation*}
	J(z, \theta) = \left ( \begin{array}{cc} \frac{\partial x}{\partial z} & \frac{\partial x}{\partial \theta} \\ & \\ \frac{\partial y}{\partial z} & \frac{\partial y}{\partial \theta} \end{array} \right ) = \left ( \begin{array}{cc} \sigma_x \; \cos \theta & -\sigma_x \; z \; \sin \theta \\ \sigma_y \; \sin \theta & \sigma_y \; z \; \cos \theta \end{array} \right ) \ ,
	\end{equation*}
	and its determinant
	\begin{equation*}
	\mathrm{det}\: J(z, \theta) = \sigma_x \sigma_y z \cos^2 \theta + \sigma_x \sigma_y z \sin^2 \theta = \sigma_x \sigma_y z \, , 
	\end{equation*}
	allows for the following substitutions
	\begin{equation*}
	\left(\frac{x}{\sigma_x}\right)^2 + \left(\frac{y}{\sigma_y} \right)^2 = \left (\frac{\sigma_x \; z \; \cos \theta}{\sigma_x} \right )^2 + \left (\frac{\sigma_y \; z \; \sin \theta}{\sigma_y} \right )^2 = z^2 \, , 
	\end{equation*}
	\begin{equation*}
	dx \; dy = \mathrm{det}\: J(z, \theta) \; dz \; d\theta = \sigma_x \sigma_y z \; dz \; d\theta \, .
	\end{equation*}
	This, in turn, leads to
	\begin{equation*}
	\alpha = \frac{1}{2 \pi} \int_0^{\sqrt{\gamma}} z \; dz \; e^{\frac{-z^2}{2}} \int_0^{2 \pi} d\theta = \int_0^{\sqrt{\gamma}} z e^{\frac{-z^2}{2}} \; dz \, .
	\end{equation*}
	Substituting: $s = -z^2 / 2$, correspondingly $ds = - z \; dz$, yields
	\begin{equation*}
	\int_0^{\sqrt{\gamma}} z e^{\frac{-z^2}{2}} \; dz = \int_{-\frac{\gamma}{2}}^{0} e^{s} ds  = \left (1 - e^{\frac{-\gamma}{2}} \right ) \, .
	\end{equation*}
	Then,
	\begin{eqnarray*}
		\alpha & = & 1 - e^{\frac{-\gamma}{2}} \\
		\hidewidth{{}\Rightarrow{}\qquad}\gamma & = & - 2 \ln \left(1 - \alpha \right) \, .
	\end{eqnarray*}

\end{document}